\documentclass[10pt, conference]{IEEEtran}
\IEEEoverridecommandlockouts
\usepackage{cite}
\usepackage{amsmath,amssymb,amsfonts}
\usepackage{algorithmic}
\usepackage{graphicx}
\usepackage{textcomp}
\usepackage{xcolor}
\usepackage{booktabs}
\usepackage{supertabular}
\usepackage{dirtytalk}
\usepackage[colorlinks,linkcolor=red,anchorcolor=green,citecolor=blue]{hyperref}
\usepackage{threeparttable}
\usepackage{multicol}
\usepackage{multirow}
\usepackage{tikz}

\def\BibTeX{{\rm B\kern-.05em{\sc i\kern-.025em b}\kern-.08em
    T\kern-.1667em\lower.7ex\hbox{E}\kern-.125emX}}
\newcommand\copyrighttext{%
  \footnotesize \textcopyright 2025 IEEE. Personal use of this material is permitted.  Permission from IEEE must be obtained for all other uses, in any current or future media, including reprinting/republishing this material for advertising or promotional purposes, creating new collective works, for resale or redistribution to servers or lists, or reuse of any copyrighted component of this work in other works.
 
  Accepted for publications at ISLPED 2025 IEEE/ACM International Symposium on Low Power Electronics and Design.}

\newcommand{\copyrightnotice}{%
\begin{tikzpicture}[remember picture,overlay]
\node[anchor=south,yshift=10pt] at (current page.south) {\fbox{\parbox{\dimexpr\textwidth-\fboxsep-\fboxrule\relax}{\copyrighttext}}};
\end{tikzpicture}%
}
\begin{document}

\newcommand{\TODO}[1]{\textcolor{red}{#1}}
\newcommand{\stef}[1]{\textcolor{black}{#1}}
\newcommand{\stefnew}[1]{\textcolor{black}{#1}}
\newcommand{\chao}[1]{\textcolor{black}{#1}}
\newcommand{\chaoRoundTwo}[1]{\textcolor{black}{#1}}
\newcommand{\chaoRoundThree}[1]{\textcolor{black}{#1}}
\newcommand{\chaoRoundFour}[1]{\textcolor{black}{#1}}
\newcommand{\mv}[1]{\textcolor{black}{#1}}
\newcommand{\xyi}[1]{\textcolor{black}{#1}}
\newcommand{\nsm}[1]{\textcolor{black}{#1}}
\newcommand{\stefnewnew}[1]{\textcolor{black}{#1}}
\newcommand{\stefFinal}[1]{\textcolor{black}{#1}}

\newcommand{\stefPostSubm}[1]{\textcolor{teal}{#1}}

\newcommand{\stefRVSD}[1]{\textcolor{black}{#1}}
\newcommand{\chaoRVSD}[1]{\textcolor{black}{#1}}

\title{Efficient Precision-Scalable Hardware for Microscaling (MX) Processing in Robotics Learning}
\author{
    Stef Cuyckens$^\dagger$, Xiaoling Yi$^\dagger$, Nitish Satya Murthy$^\dagger$, Chao Fang$^{\dagger,\ddagger,*}$, Marian Verhelst$^\dagger$ \\
	\IEEEauthorblockA{
		$^\dagger$KU Leuven, Belgium~~~~~~~~$^\ddagger$Nanjing University, China
    }
    \IEEEauthorblockA{
		Email: \{stef.cuyckens, xiaoling.yi, nitish.satyamurthy, chao.fang, marian.verhelst\}@esat.kuleuven.be
    }
}%
\maketitle
\copyrightnotice %
\vspace{-0.4cm}

\maketitle
\renewcommand{\thefootnote}{}
\footnotetext{$^*$Corresponding author. This project has been partly funded by the European Research Council (ERC) under grant agreement No. 101088865, the European Union’s Horizon 2020 program under grant agreement No. 101070374, the Flanders AI Research Program, Research Foundation Flanders (FWO) under grant No. 1S37125N, and KU Leuven.}

\begin{abstract}

\chao{\mv{Autonomous} robots require efficient on-device learning to adapt to new environments without cloud dependency. For this edge training, Microscaling (MX) data types offer a promising solution by combining integer and floating-point representations with shared exponents, reducing energy consumption while maintaining accuracy.
\chaoRoundThree{However, the state-of-the-art continuous learning processor, namely Dacapo, faces limitations with its MXINT-only support and inefficient 
vector-based grouping during backpropagation.}
\chaoRoundThree{In this paper, we present, to the best of our knowledge, the first work that addresses these limitations with two key innovations:}
}\stefFinal{(1) a precision-scalable arithmetic
unit that supports all six MX data types by exploiting sub-word parallelism and unified integer
and floating-point processing; and (2) support for square shared exponent groups to enable efficient weight handling during backpropagation, removing storage redundancy and quantization overhead.}%

\stefnewnew{We evaluate our design against Dacapo under iso-peak-throughput on four robotics workloads
\chaoRoundThree{in TSMC 16nm FinFET technology at
\chaoRVSD{400MHz},}
reaching a 
51\% lower memory footprint\stefFinal{, and 4$\times$ higher effective training throughput,
while achieving comparable energy efficiency,
enabling efficient robotics continual learning at the edge.}}

\end{abstract}

\section{Introduction}

\chao{\mv{Autonomous} robots require adaptive learning capabilities in tasks like navigation, disaster response, and human-robot collaboration when encountering unfamiliar environments~\cite{liu2021lifelong, wolczyk2021continual, pique2022controlling}. 
Traditional cloud-based learning demands continuous network connectivity while compromising data privacy and real-time responsiveness during adaptation to unexpected situations~\cite{huang2022edge, risso2023precision}. 
Instead, edge training solves these challenges by enabling robots to \mv{autonomously} perceive, learn, and modify behaviors in real-time through continuous on-device learning~\cite{zhu2024device, huang2024precision, ogbogu2023energy, Ekya_gpu, gpu2}.}

\chao{%
\chaoRoundTwo{Modern robotic systems at the edge, which rely increasingly on neural networks (NNs) for learning capabilities~\cite{fu2025dectrain, li2025deep,sim2020spartann,chua2018deepreinforcementlearninghandful}, confront a critical performance challenge.
While inference tasks can operate efficiently with integer formats like 8-bit integer (INT), training NN models requires floating-point (FP) precision to accurately capture gradient updates~\cite{rusci2023device,fang2023efficient,lu2020evaluations}.
This creates a fundamental tension in edge robotics: the need for FP precision versus the energy constraints requiring minimal bit-width.
The recently proposed Microscaling (MX) formats~\cite{rouhani2023ocp, rouhani2023microscalingdataformatsdeep},
\chaoRoundThree{which have gained significant attention from major commercial companies in Open Compute Project~\cite{rouhani2023ocp},}
provide a promising solution by combining INT and FP characteristics. 
As detailed in Table~\ref{tab:MX}, MX represents data with shared exponents across 32-element groups, offering implementations ranging from MXINT8 (pure integer) to various MXFP variants (E5M2, E4M3, E3M4, E2M5, E2M1) with different exponent-mantissa configurations.}
These formats achieve higher \mv{training} accuracy than \stef{narrow} integer-only quantization while maintaining lower memory requirements compared to full FP formats~\cite{sharify2024post,MX_implementations_single_precision,tseng2025training,chen2025oscillation}, which is critical for \mv{edge} robotic systems adapting to new environments with limited resources.}
\chao{The diversity of MX formats offers a precision-accuracy tradeoff, where lower precisions reduce energy consumption per operation.}

\begin{figure}
    \centering
    \includegraphics[width=1\linewidth]{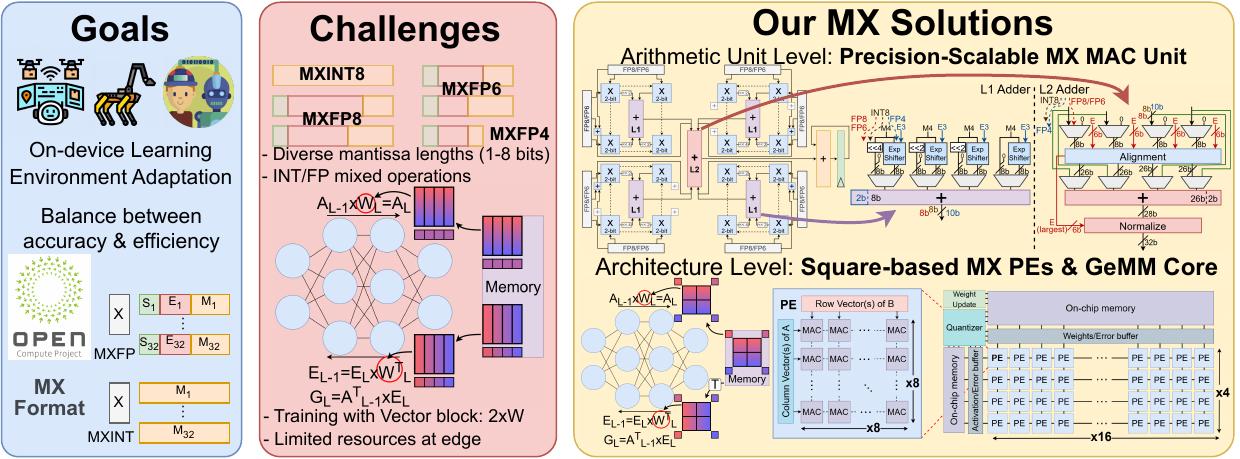}
    \caption{Overview of goals, challenges, and our contributions: (1) precision-scalable MX MAC unit supporting all six MX formats combining INT and FP operations; (2) 64-element square MX groups for training at the edge and the design of a square-based MX PE Array for GeMM Core. }%
    \label{fig:fig1}
    \vspace{-1.5em}
\end{figure}

\chao{However, \chaoRoundTwo{as shown in Fig.~\ref{fig:fig1}}, efficiently \mv{supporting all \stefnewnew{6} MX datatype variants in the robotics processing hardware} %
presents challenges at both the arithmetic unit and architecture levels.} 

\chaoRoundThree{At the arithmetic unit level, for example, the state-of-the-art (SotA) MX continuous learning processor, Dacapo~\cite{kim2024dacapo}, implements a precursor~\cite{microscalin_with_little_shifting} to the standardized MX formats~\cite{rouhani2023microscalingdataformatsdeep}, but only supports MXINT without MXFP capabilities.
This limitation restricts compatibility with the new-generation MX standard~\cite{rouhani2023microscalingdataformatsdeep},
while hampering training efficiency for workloads that would benefit from FP precision.
Beyond Dacapo, other current MX hardware implementations \cite{MX_implementations_single_precision, lin2024qservew4a8kv4quantizationcodesign} either a.) lack precision-scalable capabilities needed for \mv{the} diverse robotic \mv{training} applications, or b.) split the datapath in 1-bit processing units \cite{tahmasebi2024flexibitfullyflexibleprecision}, which leads to energy- or area-inefficient designs.}
\stef{The varied mantissa lengths \mv{of} 1, 2, 3, and 8 bits across MX data types \stef{combined with the mixture of integer and floating point formats}, create significant design complexity when supporting all formats efficiently. 
\mv{Moreover, all} existing solutions implement exact addition \cite{MX_implementations_single_precision, zhou2024}, \mv{which becomes extremely} inefficient for the \mv{two} MXFP8 format. Due to \mv{its} large exponent, MXFP8's E5M2 requires 78-bit addition in \cite{MX_implementations_single_precision}.
Additionally, transitioning between integer and floating-point operations introduces further complexity~\cite{kim2024dacapo, fang2025anda, liu2024inspire}, as different arithmetic behaviors must be accommodated within the same hardware framework.}

\stef{At the architecture level, the vector-based grouping of MX format~\cite{rouhani2023ocp} creates inefficiencies in NN training hardware accelerators. 
\chaoRoundThree{This approach, also employed by the SotA Dacapo~\cite{kim2024dacapo}, is particularly problematic during backpropagation, which requires weight matrices in both original and transposed forms for row-wise and column-wise dot-products, as presented in Fig.~\ref{fig:fig1}.}
\mv{Neither row- nor column-wise organization of the shared exponent groups is efficient for all training phases,} forcing architectural designs to a.) either maintain two separately quantized weight versions with different shared exponents \mv{as in  \cite{flexblock,FAST}}, doubling storage requirements; or b.) store unquantized weights and perform repeated quantization, increasing computation overhead and energy consumption as in \cite{Nitish_square_blocks}.} \mv{Both solutions are, however, detrimental for resource-scarce edge processing efficiency.} %

\chao{To address \mv{aforementioned challenges, this work\chaoRoundThree{, to the best of our knowledge, proposes the first}} %
efficient precision-scalable multiply-accumulate (MAC) units that support all six MX data types, \mv{embedded in a processing array optimized for square-based MX grouping, to solve all} %
storage inefficiencies of prior MX implementations.
Our key contributions include:}
\begin{itemize}
    \item \chaoRoundTwo{At the arithmetic unit level, a precision-scalable MAC that efficiently processes all MX formats using 2-bit multipliers as fundamental computational elements, unifying integer and floating-point operations within a single design while repurposing extra multipliers from precision-scaling for partial result addition.}
    \item \chaoRoundThree{At the architecture level, a novel square-based MX processing element (PE) array and learning-enabled MX general matrix multiplication (GeMM) core that replaces traditional 32-element vector-based grouping with 64-element (8$\times$8) square blocks, enabling symmetric operations across training passes without requantization while significantly reducing weight storage.}
    \item Comprehensive evaluation of our implementation against Dacapo~\cite{kim2024dacapo}, demonstrating 51\% memory footprint reduction as well as 
    4$\times$ higher effective training throughput
    for similar energy-efficiency. 
    Both systems were evaluated in TSMC 16nm FinFET at 
    \chaoRVSD{400MHz}
    for iso-peak-throughput.
\end{itemize}

\begin{table}[]
\centering
\caption{Concrete MX formats specified by \cite{rouhani2023microscalingdataformatsdeep,rouhani2023ocp}}
\label{tab:MX}
\begin{tabular}{@{}ccccccc@{}}
\toprule
\textbf{MX name}                                                   & MXINT8  & \multicolumn{2}{c}{MXFP8}   & \multicolumn{2}{c}{MXFP6} & MXFP4 \\ \midrule
\textbf{\begin{tabular}[c]{@{}c@{}}Element \\ format\end{tabular}} & INT8    & E5M2         & E4M3         & E3M2        & E2M3        & E2M1  \\ \midrule
\textbf{No. bits}                                                  & 8       & 8            & 8            & 6           & 6           & 4     \\ \midrule
& \multicolumn{6}{c}{Block size of 32 elements. Shared exponent of 8 bits.} \\ \bottomrule
\end{tabular}
\vspace{-1em}
\end{table}

\begin{figure}
    \centering
    \includegraphics[width=1\linewidth]{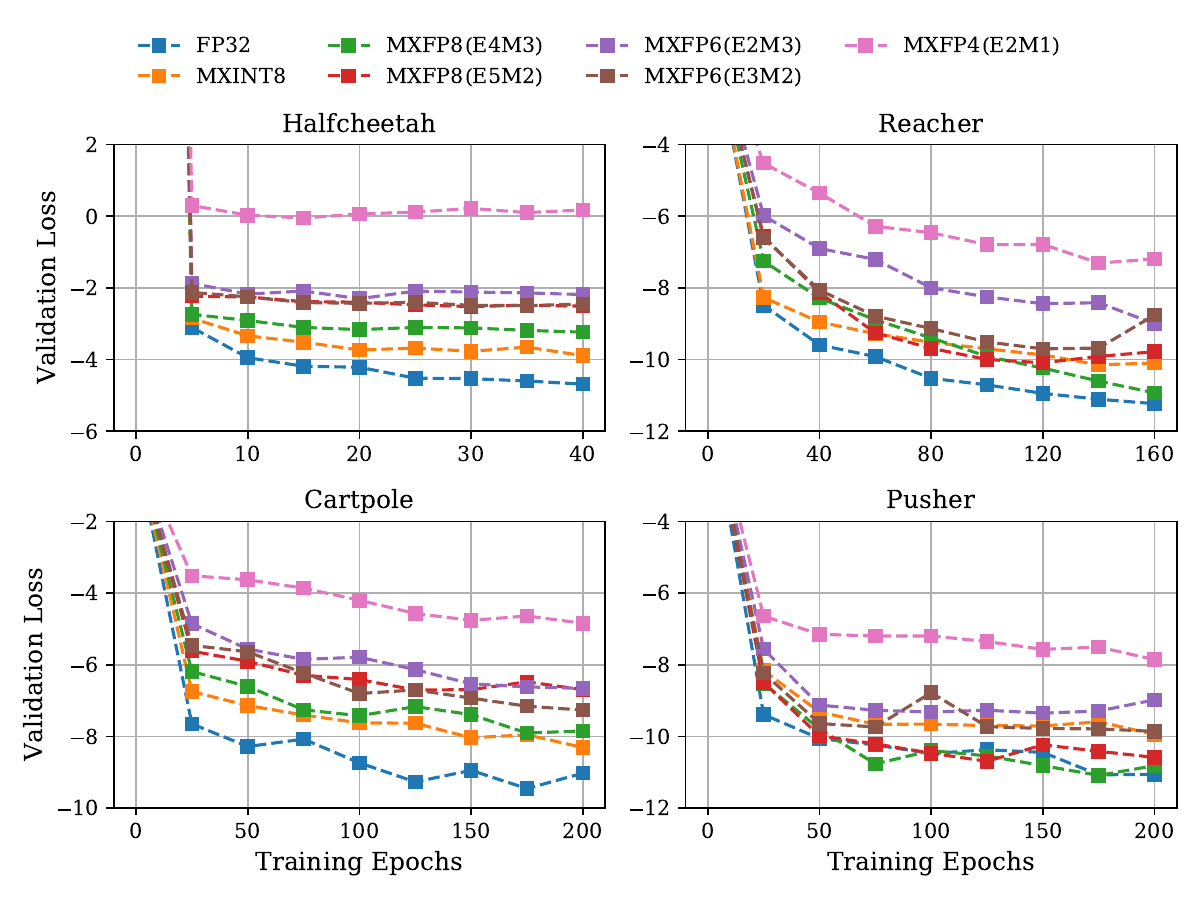}
    \vspace{-2em}
    \caption{Validation loss curves of the concrete MX data types on 4 robotic training workloads compared with the FP32 baseline, showing that FP32 can be replaced by low-bit MX types towards robotics learning.
    }
    \label{graph1}
    \vspace{-1em}
\end{figure}

\section{Background and Motivation} \label{section2}

\subsection{\chao{MX Formats}}

\chao{\chaoRoundTwo{As shown in Table~\ref{tab:MX}, }the \mv{recent} MX Standard~\cite{rouhani2023ocp} aims to achieve efficient NN implementation by minimizing accuracy loss when using 8 bits \mv{or less} to represent \mv{operands}.  %
\mv{To this end, data elements are grouped with a block size of 32, which share a joint scale denoted as X in Fig.~\ref{fig:fig1}. This scale itself is}
restricted to an \say{E8M0} format (8-bit exponent, 0-bit mantissa), which constrains it to powers-of-two. According to the standard, the scale is set to the largest power-of-two in the block, divided by the largest power-of-two representable in the element format.}

\mv{As presented in Table~\ref{tab:MX}, the data format of the individual elements within a block can be one of several}
MX-compliant formats, %
including MXFP8 E5M2, MXFP8 E4M3, MXFP6 E3M2, MXFP6 E2M3, and MXFP4 E2M1. In this notation, the first number indicates the total bit width, followed by \mv{element-specific} exponent bits and mantissa bits. For example, MXFP8 E5M2 allocates 5 bits to the exponent and 2 bits to the mantissa, \mv{together with a sign bit forming} an 8-bit representation with a wider dynamic range at the expense of precision.

\subsection{\chao{Opportunities of MX in Robotics Learning}}
\label{opportunity_mx}
\chao{Robotics increasingly demands on-device learning capabilities to adapt to dynamic environments without cloud connectivity~\cite{huang2022edge, risso2023precision, zhu2024device, huang2024precision, ogbogu2023energy}. 
For these constrained scenarios, MX formats offer an excellent alternative to high-precision formats like FP32/FP16/BF16, providing significantly reduced memory footprint and energy consumption while maintaining sufficient accuracy for gradient-based learning. The unique block-based exponent sharing scheme allows MX to capture the dynamic range necessary for representing gradients while using substantially fewer bits overall~\cite{sharify2024post,MX_implementations_single_precision,tseng2025training,chen2025oscillation}.}

\begin{figure}
    \centering
    \includegraphics[width=0.88\linewidth]{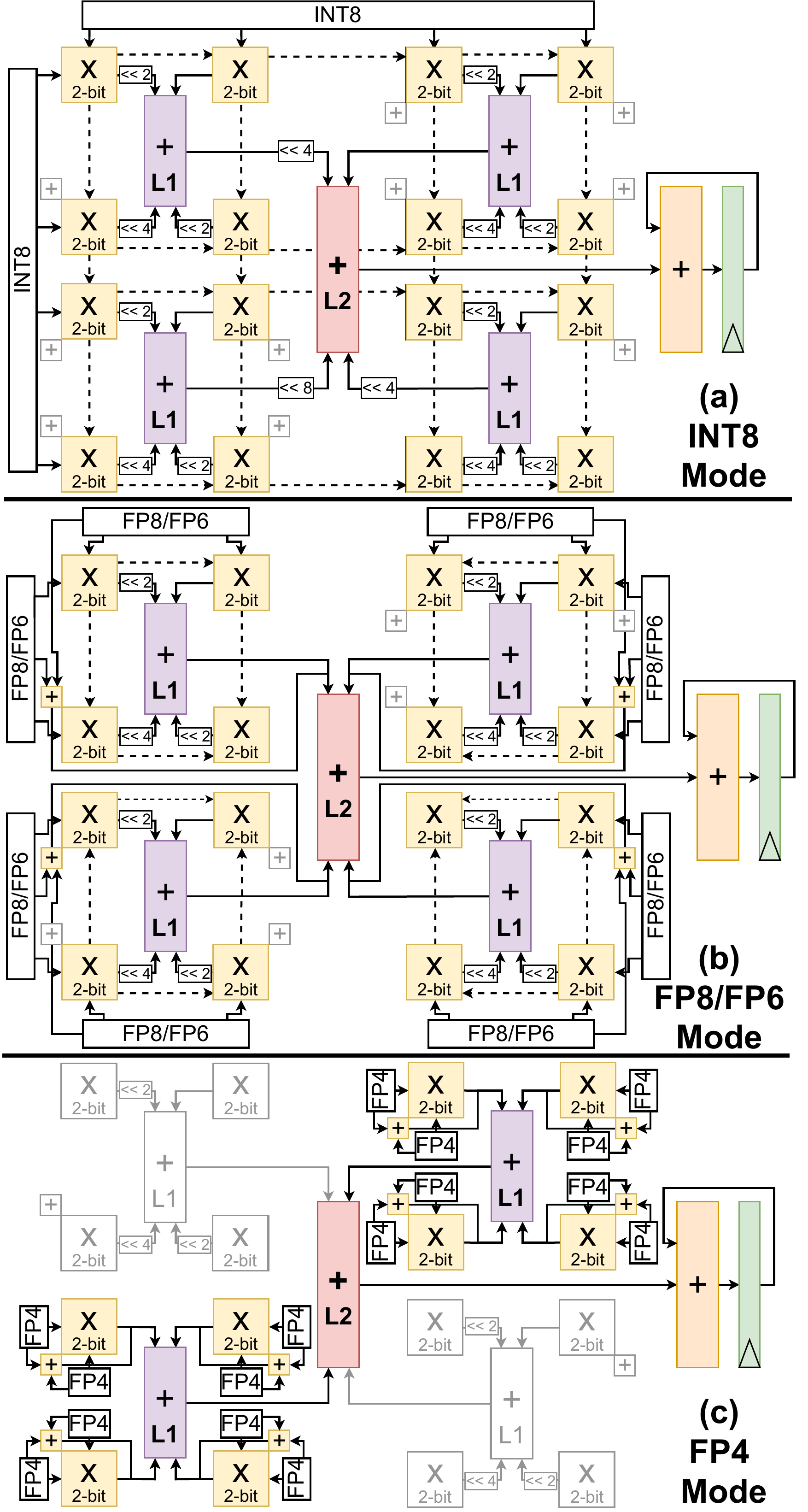}
    \caption{The precision-scalable MAC unit in the 3 scaling modes: (a) INT8, (b) FP8/FP6, and (c) FP4, respectively. \stefFinal{The multiplication is indicated in yellow, the L1 adder in purple, the L2 adder in red, the FP accumulation addition in orange, and the accumulation register in green.}}
    \label{fig:ST_mac}
    \vspace{-1.5em}
\end{figure}

Fig.~\ref{graph1} summarizes our experimental evaluation of MX processing in robotics learning \cite{chua2018deepreinforcementlearninghandful} across multiple tasks \mv{and MX precision settings.} \stefnewnew{These tasks involve training NNs to predict the system dynamics for continuous control robotics systems from \cite{chua2018deepreinforcementlearninghandful}.} \mv{The experiments} reveal a critical finding: different robotic learning workloads benefit from different precision configurations. 
\stefnew{Specifically, MXFP8 E4M3 demonstrates superior efficiency in training pusher and reacher tasks, which are relevant for robot-object interaction~\cite{pique2022controlling}. Conversely, MXINT8 achieves the highest performance in cartpole and halfcheetah workloads, which represent balancing tasks~\cite{wang2021balance}.}

\section{Arithmetic Unit Innovations} \label{section3}

\subsection{\chaoRoundTwo{Precision-Scalable MX MAC Unit}}

\chaoRoundTwo{Robotics learning demands efficient precision-scalable compute to balance training throughput and inference efficiency.
Our precision-scalable MX MAC unit meets this need by supporting all MX data formats, as shown in Table~\ref{tab:MX} with format-specific optimizations.}

\chaoRoundTwo{Fig.~\ref{fig:ST_mac} presents the architecture of our precision-scalable MX MAC unit across its three operating modes. 
\mv{The architecture of this flexible MAC unit is composed of 16 elementary} 2-bit multiplication units, \mv{which are flexibly interconnected to} %
efficiently \mv{support} all \mv{MX} formats, while reducing adder overhead compared to \mv{designs composed of} 1-bit \mv{processing blocks}~\cite{tahmasebi2024flexibitfullyflexibleprecision}.
Our architecture maximizes hardware reuse across precision levels, while maintaining optimal performance for each mode.}

\chaoRoundTwo{The MAC flexibly operates in three distinct \mv{operating} modes: INT8, FP8/FP6, and FP4, \mv{with} dynamic hardware reconfiguration across these precision modes.
As shown in Fig.~\ref{fig:ST_mac}(a), in INT8 mode, all sixteen 2-bit multipliers work together to compute a single 16-bit result \mv{of the INT8$\times$INT8 multiplication, in which} the exponent adders remain inactive.}

For the FP8/FP6 mode, illustrated in Fig.~\ref{fig:ST_mac}(b), the same hardware processes four parallel \mv{FP8/6$\times$FP8/6} products simultaneously, with each product utilizing four 2-bit multipliers for mantissa multiplication and \stefnewnew{one dedicated 5-bit exponent adder.}%

\mv{Finally, in the FP4 mode of} Fig.~\ref{fig:ST_mac}(c), each active 2-bit multiplier is paired with a 2-bit exponent adder to form a complete FP4 processing unit. \mv{This mode does not utilize its full computational parallelism, and limits parallelism to eight parallel FP4$\times$FP4} results due to bandwidth limitations. In this configuration, %
the bandwidth (BW) is kept equal to the FP8/6 mode, while \mv{still 8 results are computed in parallel to be added up in one} 8$\times$8 shared exponent block. %
\mv{This design choice and its array level implications are discussed in more detail} in %
Sec.~\ref{section4}. %

This configuration enables consistent throughput across precision levels, with computational density increasing at lower precisions.

\chaoRoundTwo{To efficiently \mv{sum-}reduce multiple parallel products in lower precision modes, we develop a hierarchical MX accumulator with a two-level adder architecture.
This approach implements a Sum-Together scheme~\cite{MAC_lvl_analysis} that repurposes internal adders based on the \mv{selected precision} mode, \mv{accumulating} partial products in INT8 mode and \mv{adding up} parallel FP results in FP8/FP6 and FP4 modes.
By generating a single output \mv{per MAC processing unit} regardless of precision, we simplify the \mv{array level} datapath~\cite{Linyan}, while efficiently managing mixed-precision accumulation. \mv{Next, we will} elaborate on the hierarchical \mv{two-level} accumulator design and its critical path optimizations.}

\begin{figure}
    \centering
    \includegraphics[width=\linewidth]{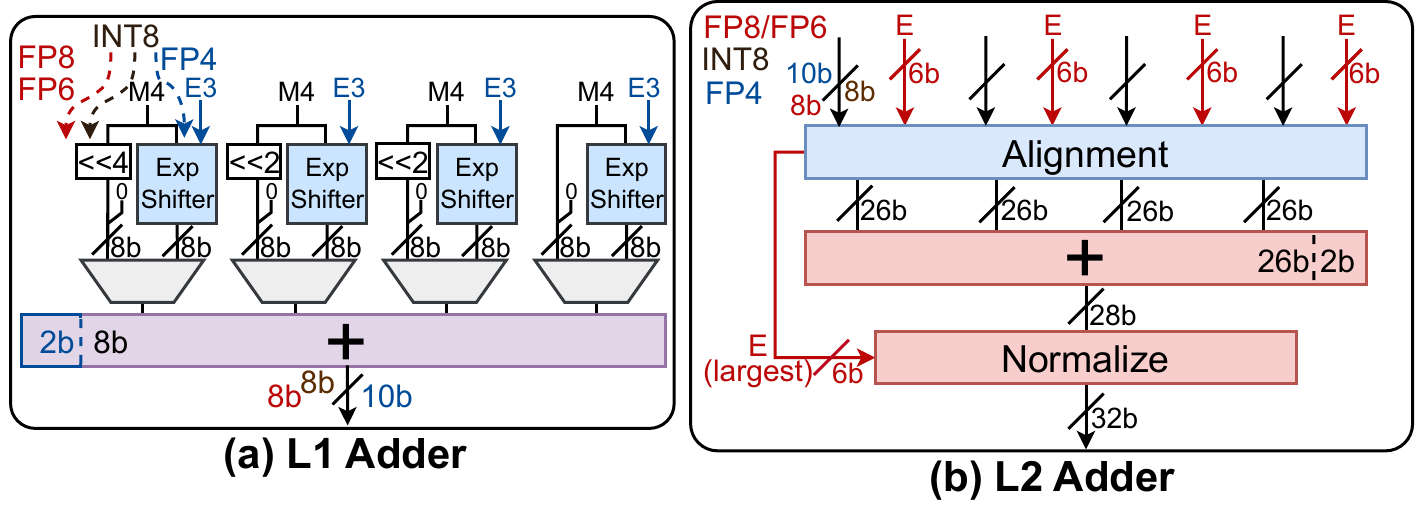}
    \vspace{-2em}
    \caption{\stefRVSD{The L1 and L2 adders, the INT8 path is indicated in brown, FP8/FP6 in red and FP4 in blue.}}%
    \vspace{-1em}
    \label{fig:accum_path}
\end{figure}

\begin{figure}[t]
    \centering
    \includegraphics[width=1\linewidth]{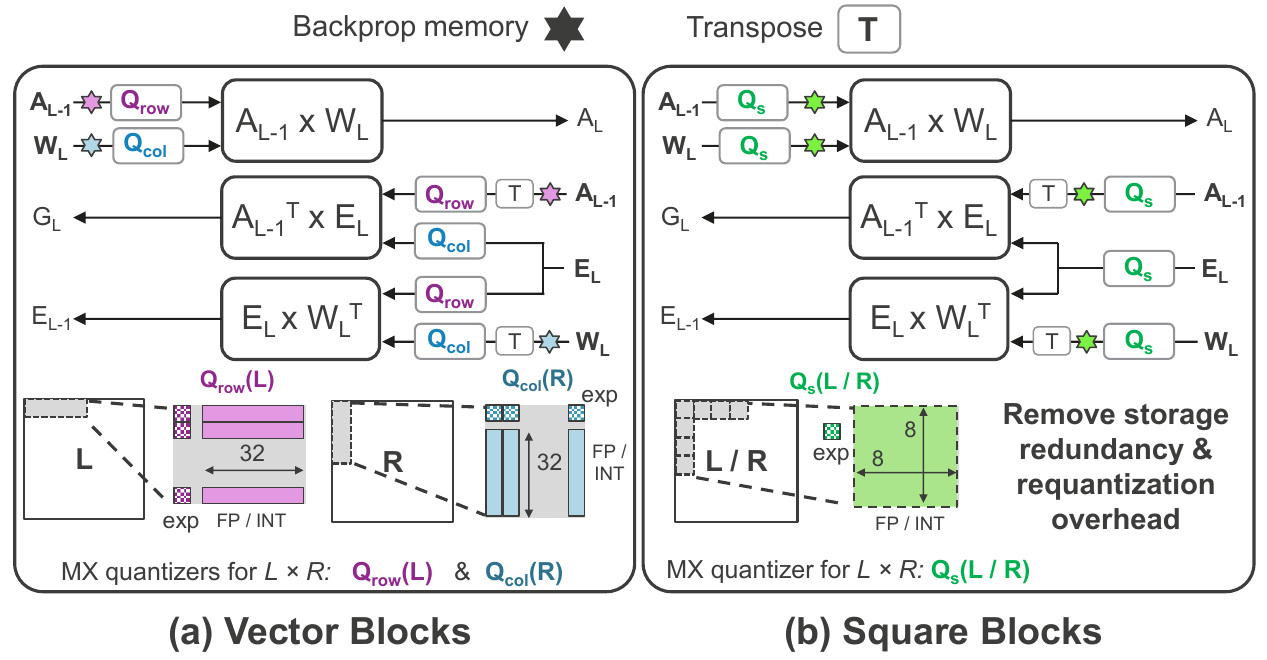}
    \caption{\nsm{Training computation graphs with MX vector and square block quantizers. Computational and memory footprint gains can be achieved for square blocks, by enabling storage of block-quantized parameters for backpropagation.}} %
    \label{mx_square_vector_blocks}
    \vspace{-1.5em}
\end{figure}

\subsection{\chaoRoundTwo{Area-Efficient Hierarchical MX Accumulator}}

\chaoRoundTwo{As shown in Fig.~\ref{fig:accum_path}, the hierarchical accumulator architecture employs Level-1 (L1) and Level-2 (L2) adders that efficiently handle different MX precision modes while maximizing hardware reuse. This design addresses three key challenges: managing various precision-dependent addition operations, guaranteeing computational accuracy, and balancing critical paths across precision modes.}

\chaoRoundTwo{\textbf{L1 Adder.} \stefnew{It is illustrated in Fig.~\ref{fig:accum_path}(a). For INT8/FP8/FP6 operations, the L1 adder accumulates 4-bit integer partial products with appropriate shifts, as shown in \mv{Fig.~\ref{fig:ST_mac}(a) and (b)}, to form an 8-bit output. For FP4 operations, it sums four different completed E3M4 multiplication results by directly shifting mantissas by their exponents, which avoids the need to find the maximum exponent first. This simplification leverages the limited range of E3M4 exponents (0-4) and enables significant hardware reuse, as the same integer adder serves all precision modes, \stefnew{with a 2-bit extension for FP4 mode.}}}

\chaoRoundTwo{\textbf{L2 Adder.} As presented in Fig.~\ref{fig:accum_path}(b), the L2 adder faces different requirements across precision modes. While a direct shifting approach for \mv{INT accumulation in} FP8/FP6 mode would require an impractical 78-bit integer adder due to their large exponent ranges (±30 for FP8 E5M2), our implementation uses FP32 addition with a more modest 26-bit mantissa adder.
To address the prevalence of subnormal numbers in narrow FP formats, we extend the internal mantissa bitwidth by only 2 bits rather than implementing costly normalization circuitry. This solution guarantees FP32-accurate addition even with non-normalized inputs while minimizing hardware overhead.}

\textbf{Critical Path Optimizations.} We implement precision-specific bypasses to balance delay paths across different modes. Both the INT8 mode and the FP4 mode have a longer critical path in the L1 adder, INT8 due to its sign-magnitude conversion, and FP4 due to the required variable shifters. However, both modes can bypass the alignment logic in the L2 adder, which is only needed for FP8/6. This mode-specific bypassing strategy balances the critical path lengths across modes and \stefRVSD{is implemented by the synthesis tools.}

\section{Architectural Innovations} \label{section4}

\subsection{\chaoRoundThree{Square-based MX PE Array}} \label{section4b}

\chaoRoundTwo{Traditional MX architecture employs 32-element vector blocks~\cite{rouhani2023microscalingdataformatsdeep}, which creates significant inefficiencies during NN training. 
\nsm{As shown in Fig.~\ref{mx_square_vector_blocks}(a), a training graph involves GeMM computations with operands of  activations, weights, and backpropagation errors in its original and transposed form. Since, performing the same vector quantization -- row or column blocks on the original and transposed matrices produce different results, there is a need of maintaining two separate versions. This doubles memory requirements, leading to a critical challenge for resource-limited edge devices~\cite{weights_big,tinycare}.}}

\chaoRoundTwo{We propose 64-element square blocks ($8 \times 8$) with shared exponents to address these limitations. 
As illustrated in Fig.~\ref{mx_square_vector_blocks}(b), our square blocks enable symmetric operations, allowing the same quantized representation to serve both passes without requantization. %
The implementation uses a shared exponent for all 64 elements within each block, enabling efficient scaling across different precision requirements. We select the $8 \times 8$ configuration to balance granularity and efficiency while maintaining compatibility with \mv{the MX standard defining groups as multiples of 32 elements. Here two 32-element blocks share the same exponent.}} %

\stefnew{Our design processes these square blocks through \mv{the PE Array}, illustrated in Fig.~\ref{block_arch}, which consist of 64 \mv{of the proposed flexible} MAC units, each supporting the multiple precision modes (INT8, FP8/FP6, FP4). \chaoRoundThree{GeMM} of two $8 \times 8$ blocks requires adding up 8 multiplication results for each of the 64 outputs. Our PE \mv{Array} handles a block multiplication in 8 \mv{subsequent clock} cycles in INT8 mode, in 2 cycles in FP8/FP6 mode, and in 1 cycle in FP4 mode, by using the precision-scalable MAC units from Sec.~\ref{section3} to increase throughput for the FP modes. Fig.~\ref{block_arch} shows this throughput increase by indicating which parts of the input blocks are processed in one cycle for each precision mode.}

\stefnew{\mv{Due to its usage of square blocks with a shared exponent, this} architecture \mv{avoids costly requantizations in the back-propagation step and reduces weight storage by nearly 50\% compared to the SotA MX training hardware Dacapo~\cite{kim2024dacapo} while maintaining computational efficiency across training stages}. This makes it ideal for on-device learning applications where memory constraints are particularly challenging.}

\subsection{\chaoRoundTwo{Learning-enabled MX GeMM Core}}

\begin{figure}[t]
    \centering
    \includegraphics[width=1\linewidth]{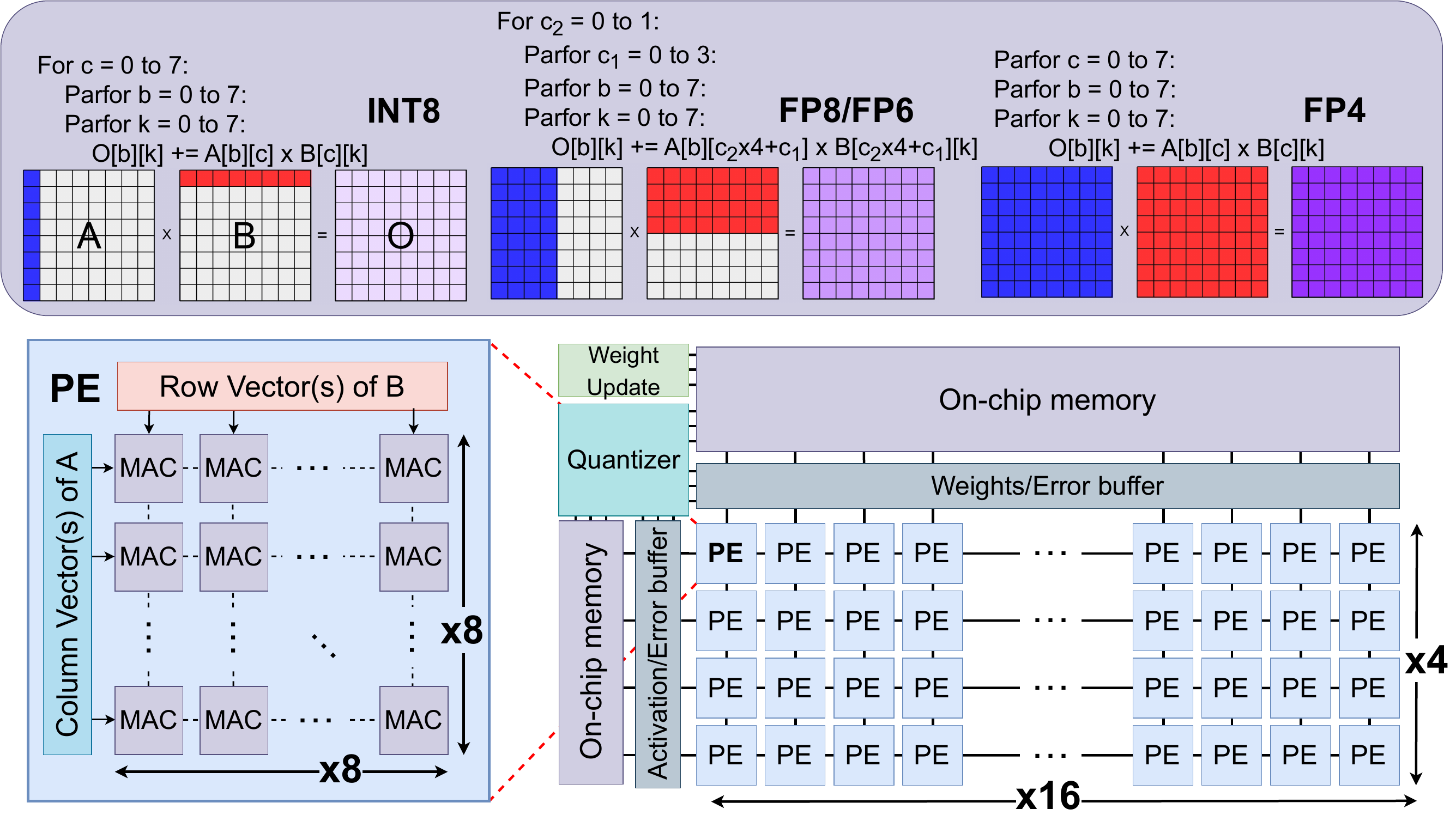}
    \caption{The PE \mv{Array} that handles the multiplication of two 64-element square blocks, for INT8 this is executed in 8 cycles, for FP8/FP6 this takes 2 cycles, and for FP4 in 1 cycle. The PE Array is part of our GeMM core of 4x16 PE Arrays.}
    \label{block_arch}
    \vspace{-1em}
\end{figure}

\chaoRoundTwo{The \mv{64-MAC} MX PE \mv{Arrays} are arranged in a $4 \times 16$ grid to form our learning-enabled MX GeMM Core, as illustrated in Fig.~\ref{block_arch}. 
This \mv{grid} configuration is optimized for edge device training constraints, with dimensions carefully selected based on typical workload characteristics. 
The height dimension of 4 corresponds to a batch size of 32 divided by our $8 \times 8$ square dimension, while the width of 16 achieves an optimal balance between bandwidth efficiency and computational parallelism. 
This rectangular configuration addresses the memory limitations inherent to edge devices, where smaller batch sizes are \mv{preferred} as activations must be stored during backward passes.}

\chaoRoundTwo{The MX GeMM Core implements an output-stationary dataflow where partial results accumulate locally within each PE to minimize FP32 intermediate data transfers.} This requires the shared exponents to be integrated into each arithmetic unit's FP accumulator for proper alignment. \mv{To this end, the} shared exponents of the input blocks are added together at PE-level \mv{and subsequently applied} to the output of the L2 adder of each MAC. 

\chaoRoundTwo{The MX GeMM Core exhibits distinct execution patterns across training stages: During the forward pass, the \mv{grid} processes matrix multiplications with high utilization. The backward pass essentially mirrors the forward pass in reverse order with similar utilization patterns. The weight gradient calculation stage presents unique challenges as accumulation occurs over the smaller batch dimension of 32, necessitating more frequent memory writes and substantially reducing array utilization.}
\chaoRoundTwo{The GeMM Core operates with a maximum \mv{memory} bandwidth of 5280 bits/cycle (approximately \stefRVSD{264}GB/s), fully utilized during FP8 and FP4 operations. During stall cycles, particularly in the weight gradient calculation stage, this bandwidth is dedicated to writing back the FP32 outputs, which are then transferred to the quantizer as shown in Fig.~\ref{block_arch}. For weight updates, the actual addition of gradients to weights occurs in a separate accelerator block, enabling our GeMM Core to focus on matrix multiplication operations while maintaining efficient pipelining on resource-constrained edge devices.}

\section{Experimental Results} 
This section presents the evaluation of the proposed \mv{MX scalable MAC unit} introduced in Sec.~\ref{section3}, followed by an assessment of the \mv{PE Array} with shared exponent support described in Sec.~\ref{section4b}. Finally, the performance of our learning-enabled MX GeMM core is compared with \mv{the Dacapo processor, a SotA MX continuous learning processor \cite{kim2024dacapo}}. All experimental results are obtained from synthesis in a TSMC 16nm FinFET technology. \mv{Switching annotated} power and energy analyses are obtained with Synopsys PrimeTime PX.

\begin{table}[t]
\centering
\caption{\stefRVSD{Comparison of implementation variants of the Precision-scalable MX MAC}}
\label{tab:MX_MAC_comparison}
\resizebox{0.5\textwidth}{!}{%
\begin{tabular}{lcc ccccccc}
\toprule
\multirow{3}{*}{MX MAC} & \multirow{3}{*}{\begin{tabular}{c}Freq.\\{}[MHz]\end{tabular}} & \multirow{3}{*}{\begin{tabular}{c}Area\\{}[$\mu m^2$]\end{tabular}} & \multicolumn{6}{c}{Energy [pJ/OP]} \\
& & & INT8 & MXFP8 & MXFP8 & MXFP6 & MXFP6 & MXFP4 \\
& & & & (E5M2) & (E4M3) & (E3M2) & (E2M3) & (E2M1) \\
\midrule
mant. add. ext. & 400 & 2341.35 & 1.62 & 1.18 & 1.22 & 1.16 & 1.22 & 0.29 \\
norm. inputs & 400 & 3640.78 & 2.46 & 2.41 & 2.58 & 2.38 & 2.58 & 0.60 \\
mant. add. ext. + bypass & 400 & 2078.42 & 1.36 & 1.14 & 1.16 & 1.11 & 1.16 & 0.28 \\
\bottomrule
\end{tabular}
}
\vspace{-1em}
\end{table}

\subsection{\chaoRoundThree{Evaluation of \mv{Precision-scalable MX MAC unit}}}%

\stefnew{
This section presents an analysis of \mv{the different design choices} %
of the precision-scalable MX MAC, as introduced in Sec. \ref{section3}. %
Each variant was synthesized independently, %
and the average energy per multiplication operation (OP) across 500 cycles using randomly generated input data \mv{was extracted}. This evaluation aims to quantify the benefits of the proposed arithmetic unit optimizations described in Sec.~\ref{section3}. Table~\ref{tab:MX_MAC_comparison} evaluates the two \mv{proposed} methods for handling non-normalized inputs at L2: (i) extending the mantissa adder by two bits or (ii) normalizing the inputs at L2. \mv{The results clearly show} the merits of extending the mantissa adder width compared to a normalization operation for each input.}
Additionally, the impact of incorporating bypass logic to balance the critical paths across different modes is examined. The findings demonstrate a clear advantage, as delay optimizations can be minimized on the larger FP8/FP6 path within the L2 adder, leading to a reduction in area. This substantial reduction is attributed to the fact that the critical timing can now be met much easier by the tools, requir\stefRVSD{ing} less large drive strength and buffer cells. 
This evaluation does not yet include the shared exponent logic.

\begin{figure}[t]
    \centering
    \includegraphics[width=\linewidth]{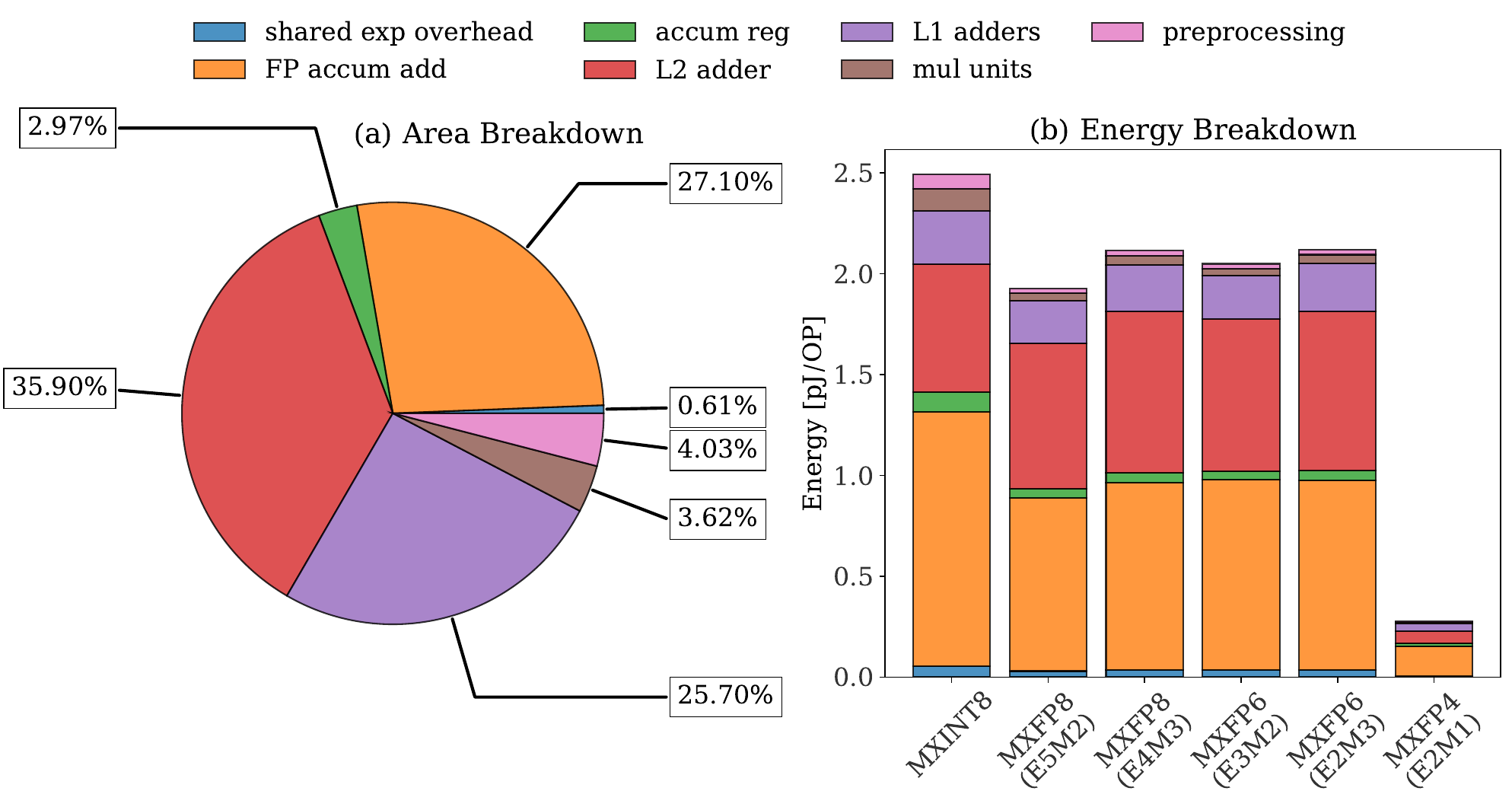}
    \vspace{-1.5em}
    \caption{\stefRVSD{The area and energy/OP (multiplication) breakdown of the PE Array.}}
    \label{energy_breakdown}
    \vspace{-1em}
\end{figure}

\begin{table}[t]
    \centering
    \caption{Memory footprint comparison of ours with Dacapo for MXINT8/MX9 and the unquantized FP32 baseline}
    \resizebox{0.5\textwidth}{!}{%
    \begin{tabular}{c|c|c c|c c c c|c}
    \toprule
         Batch & \multirow{2}{*}{Method} & \multicolumn{2}{c|}{Inference [KB]} & \multicolumn{4}{c|}{Training [KB]} & \multirow{2}{*}{Total [KB]} \\
         Size &  & W & A & $W^T$ & $A^T$ & E (row.) & E (col.) & \\
         \midrule
         \multirow{3}{*}{16} & FP32 & 576.0 & 0.0 & 0.0 & 50.0 & 16.0 & 0.0 & 642.0 (1.00$\times$) \\
          & Dacapo \cite{kim2024dacapo} & 162.0 & 4.5 & 162.0 & 14.1 & reuse A & 4.5 & 347.1 (1.85$\times$)\\
          & \textbf{Ours} & \textbf{146.3} & \textbf{0.0} & \textbf{0.0} & \textbf{12.7} & \textbf{4.1} & \textbf{0.0} & \textbf{163.1 (3.94$\times$)} \\
         \midrule
         \multirow{3}{*}{32} & FP32 & 576.0 & 0.0 & 0.0 & 100.0 & 32.0 & 0.0 & 708.0 (1.00$\times$) \\
          & Dacapo \cite{kim2024dacapo} & 162.0 & 9.0 & 162.0 & 28.1 & reuse A & 9.0 & 370.1 (1.91$\times$)\\
          & \textbf{Ours} & \textbf{146.3} & \textbf{0.0} & \textbf{0.0} & \textbf{25.4} & \textbf{8.1} & \textbf{0.0} & \textbf{179.8 (3.94$\times$)} \\
         \midrule
         \multirow{3}{*}{64} & FP32 & 576.0 & 0.0 & 0.0 & 200.0 & 64.0 & 0.0 & 840.0 (1.00$\times$) \\
          & Dacapo \cite{kim2024dacapo} & 162.0 & 18.0 & 162.0 & 56.3 & reuse A & 18.0 & 416.3 (2.02$\times$)\\
          & \textbf{Ours} & \textbf{146.3} & \textbf{0.0} & \textbf{0.0} & \textbf{50.8} & \textbf{16.3} & \textbf{0.0} & \textbf{213.4 (3.94$\times$)} \\
         \bottomrule
    \end{tabular}
    }
    \label{tab:mem_comp}
    \vspace{-1em}
\end{table}

\begin{table}
    \centering
    \caption{\stefRVSD{Comprehensive Comparison of Ours and the SotA Dacapo \cite{kim2024dacapo}}}%
    \resizebox{0.48\textwidth}{!}{%
    \begin{tabular}{c|c|c|c}
    \toprule
    \multicolumn{2}{c|}{Specifications} & Ours & Dacapo~\cite{kim2024dacapo} \\
    \midrule
    \multicolumn{2}{c|}{Freq. [MHz]} & 400 & 400 \\
    \multicolumn{2}{c|}{Area [$mm^2$]} & 8.92 & 8.52 \\
    \multicolumn{2}{c|}{Max. BW [GB/s]} & 264 & 512 \\
    \multicolumn{2}{c|}{Mem. [KB]} & 179.78 & 370.13 \\
    \multicolumn{2}{c|}{Amount of MACs} & 4096 & 4096 \\
    \midrule
    \multirow{3}{*}{E/op [pJ]} & MXINT8 vs. MX9 & 2.43 & 2.68 \\
    & MXFP8/FP6 vs. MX6 & 1.92 - 2.12 & 1.46 \\
    & MXFP4 vs. MX4 & 0.278 & 0.358 \\
    \midrule
    \multicolumn{2}{c|}{Batch Size} & 32 & 32 \\
    \midrule
    \multirow{3}{*}{Train Latency/Batch [$\mu s$]} & MXINT8 vs. MX9 & 13.575 & 50.5 \\
    & MXFP8/FP6 vs. MX6 & 6.03 & 30.7 \\
    & MXFP4 vs. MX4 & 4.7625 & 25.75 \\
    \bottomrule
    \end{tabular}
    }
    \label{tab:DACvsUS}
    \vspace{-1em}
\end{table}

\subsection{\chaoRoundThree{Evaluation of PE Array}}

This section presents a detailed analysis of the area and energy breakdown for the \mv{PE Array,} %
as illustrated in Fig.~\ref{energy_breakdown}. This analysis provides insights into the most resource-intensive components within a realistic \mv{array} implementation. The \mv{Array} uses MAC units with bypassing in the L2 adder. 
The average power for the energy breakdown was computed over 100 block multiplications with random input data. The energy was divided by a total number of 51,200 multiplication OPs for the energy breakdown as shown in Fig. \ref{energy_breakdown}. %

The energy breakdown indicates that FP accumulation addition is the most energy-intensive component of the architecture. 
Furthermore, the results show that the overhead associated with the shared exponent is negligible. %

In contrast, the area breakdown reveals that the L1 and L2 adders account for the largest portion of the total area. This asymmetry between area and energy consumption arises from the fact that the L1 and L2 adders contain mode-specific components, leading to increased area utilization. %

\subsection{\stefnew{Evaluation of GeMM Core Compared to the SotA}}%

\stefnew{Our complex MX processor core, as depicted in Fig. \ref{block_arch}, is compared with the reference MX-like SotA training processor Dacapo \cite{kim2024dacapo}. Both are synthesized with a clock frequency of \stefRVSD{400MHz} in TSMC 16nm FinFET and both processors use the same total number of precision-scalable MACs for iso-peak-throughput. The precision-scalable MAC units of Dacapo support INT8, INT4, and INT2 formats for the individual elements. They use vector blocks of 16 elements with a shared exponent of 8 bits. Furthermore, they also use a second level of shared exponent, an extra 1-bit exponent for each subgroup of 2 elements. This data format was developed in \cite{microscalin_with_little_shifting}, and is a precursor of the MX standard defined later in \cite{rouhani2023microscalingdataformatsdeep}. The data formats of Dacapo are referred to as MX9, MX6, and MX4 but hence do not comply with the more recent MX standard.}
\stefnew{Table~\ref{tab:mem_comp} compares the memory footprint of Dacapo and our GeMM accelerator against the FP32 baseline for the pusher workload from Fig.~\ref{graph1}. The pusher workload uses a fully-connected NN of 4 layers with input and output dimensions of 32 and hidden layer dimensions of 256. The vector block quantization of Dacapo requires 2.06$\times$ more memory footprint than our square block approach.}

\stefnew{To compare the energy/OP (multiplication) of our design vs. Dacapo, we evaluate one of our PE arrays vs. an 8$\times$8 systolic array of Dacapo on 51,200 multiplication OPs with randomly generated input data. The results in Table \ref{tab:DACvsUS} show that we use \stefRVSD{10\% less} energy/OP in MXINT8 mode and \stefRVSD{29\% less in MXFP4 mode}, but are \stefRVSD{more expensive in MXFP8 mode with 1.32$\times$ to 1.45$\times$} the energy of Dacapo in its corresponding precision mode.}
\stefnew{The rest of Table \ref{tab:DACvsUS} compares our processor core \stefFinal{of 4x16 PE Arrays} vs Dacapo's systolic array with the same number of MACs. Our design requires \stefRVSD{1.05$\times$ more area}, %
1.94$\times$ less BW, and 2.06$\times$ less memory. The latency of one training loop over the fully-connected NN of the pusher workload is compared, with latency data from Dacapo calculated by \cite{kim2024dacapo}.
\chaoRoundFour{The training latency of our design is significantly lower than Dacapo under iso-peak-throughput due to Dacapo's overhead from systolically shifting data in and out of its systolic array.}}

\begin{figure}[t]
    \centering
    \includegraphics[width=\linewidth]{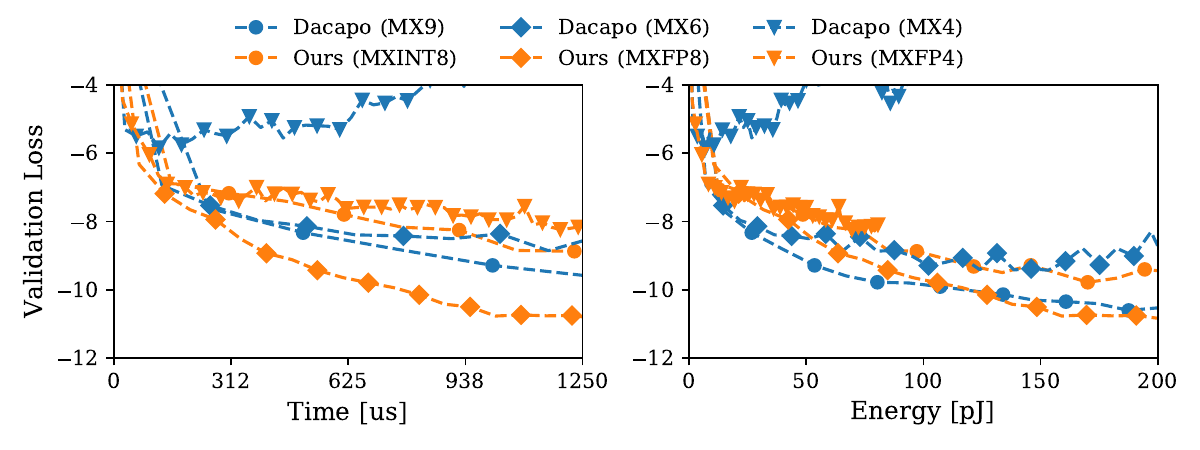}
    \vspace{-1em}
    \caption{\stefRVSD{Comparison of Dacapo and our MX GeMM core on pusher task validation loss curves given a 1250 $\mu$s time budget and 200 pJ energy budget, respectively.
    }}
    \label{fig:time_energy_to_accuracy}
    \vspace{-1em}
\end{figure}

\stefnew{We also compare the training performance of our accelerator with Dacapo. Fig. \ref{fig:time_energy_to_accuracy} (left) shows the validation loss vs training time on the respective processor for the pusher workload, which shows that MXFP8 on our accelerator is the clear winner for fast and accurate training on the pusher task. Fig. \ref{fig:time_energy_to_accuracy} (right) shows the energy-efficiency of training, where the performance of MXFP8 is comparable to MX9 because MXFP8 needs more epochs for training.}

\section{Conclusion}

\chao{This work addresses the challenges of on-device learning for autonomous robots through two key innovations for Microscaling (MX) data type processing: (1) a precision-scalable arithmetic unit that dynamically adapts to different mantissa precisions while unifying integer and floating-point processing, and (2) a 64-element \mv{PE array compatible with square shared exponent blocks, to} %
streamline weight handling during backpropagation. 
\stefFinal{Our approach increases effective training throughput by 4$\times$,} reduces memory footprint by 51\% compared to state-of-the-art MX implementations
under iso-peak-throughput on multiple robotics learning tasks, while achieving similar energy-efficiency and supporting all six MX data formats described in the MX standard.
These advances enable efficient continual learning at the edge for robotics applications, allowing autonomous robots to adapt to new environments without cloud dependency.} %

\clearpage
\bibliographystyle{IEEEtran}
\bibliography{refs}

\end{document}